\documentclass{article}

\def\*{{\bf ***}}

\def\a{\alpha}
\def\b{\beta}

\def\ga{\gamma}
\def\de{\delta}   

\def\phi{\varphi}

\def\s{\sigma}
\def\z{\zeta}
\def\om{\omega}
\def\th{\theta}
\def\vth{\vartheta}

\def\D{{\cal D}}

\def\h{{\cal H}}

\def\J{{\cal J}}

\def\N{{\cal N}}

\def\R{{\bf R}}

\def\T{{\rm T}}
\def\V{{\cal V}}
\def\W{{\cal W}}

\def\Ga{\Gamma}

\def\Om{\Omega}

\def\pa{\partial}

\def\d{{\rm d}}       
\def\w{\wedge}
\def\xb{{\bf x}}

\def\o+{\oplus}

\def\ss{\subset}
\def\sse{\subseteq}

\def\<{\langle}
\def\>{\rangle}

\def\interno{\hskip 2pt \vbox{\hbox{\vbox to .18
truecm{\vfill\hbox to .25 truecm
{\hfill\hfill}\vfill}\vrule}\hrule}\hskip 2 pt}

\def\({\left(}
\def\){\right)}
\def\[{\left[}
\def\]{\right]}
\def\=#1{\bar #1}
\def\~#1{\widetilde #1}
\def\.#1{\dot #1}
\def\^#1{\widehat #1}
\def\"#1{\ddot #1}

\def\mapright#1{\smash{\mathop{\longrightarrow}\limits^{#1}}}

\def\ref#1{\cite{#1}}

\def\Remark#1{\medskip \noindent {\bf Remark {#1}}}

\begin{document}

\title{\bf Maximal degree variational principles}

\author{Giuseppe Gaeta\footnote{Supported by ``Fondazione CARIPLO
per la ricerca scientifica''} {~}\footnote{e-mail: g.gaeta@tiscali.it} \\
{\it Dipartimento di Matematica, Universit\'a di Milano} \\
{\it via Saldini 50, I--20133 Milano (Italy)} \\ {~~} \\
Paola Morando\footnote{e-mail: morando@polito.it} \\
{\it Dipartimento di Matematica, Politecnico di Torino} \\
{\it Corso Duca degli Abruzzi 24, I--10129 Torino (Italy)} }

\date{~}

\maketitle

\noindent
{\bf Summary.} Let $M$ be smooth $n$-dimensional manifold, fibered over a $k$-di\-men\-sio\-nal submanifold $B$ as $\pi:M \to B$, and $\vth \in \Lambda^k (M)$; one can consider the functional on sections $\phi$ of the bundle $\pi$ defined by $\int_D \phi^* (\vth)$, with $D$ a domain in $B$. We show that for $k = n-2$ the variational principle based on this functional identifies a unique (up to multiplication by a smooth function) nontrivial vector field in $M$, i.e. a system of ODEs. Conversely, any vector field $X$ on $M$ satisfying $X \interno \d \vth = 0$ for some $\vth \in \Lambda^{n-2} (M)$ admits such a variational characterization. We consider the general case, and also the particular case $M = P \times \R$ where one of the variables (the time) has a distinguished role; in this case our results imply that any Liouville (volume-preserving) vector field on the phase space $P$ admits a variational principle of the kind considered here.
\medskip

\section*{Introduction}

A variational principle is defined by a functional $I_D$ over sections $\phi$ of a fiber bundle $\pi : M \to B$; this is given in terms of pullbacks $\phi^* (\vth)$ of a form $\vth$ such that $\phi^* (\vth)$ is a volume form on $B$, and defined by 
$$ I_D (\phi ) \ := \ \int_D \phi^* (\vth)  $$
where $D$ is a domain in $B$ (precise definitions for this and other notions will be given in section 1 below). 
Let $\V_D (\pi)$ denote the set of vector fields vertical for the fibration $\pi:M \to B$ and vanishing on $\pi^{-1} (\pa D)$; by a standard theorem of variational analysis, the variation of $I_D (\phi)$ under $V \in \V_D (\pi)$ is given by 
$$ (\delta_V I_D) (\phi) \ = \ \int_D \phi^* (V \interno \d \vth ) \ ;   $$
one says that $\phi$ is a {\it critical point} of $I_D$, or equivalently a {\it critical section}, if $(\de_V I_D) (\phi) = 0 $ for all $V \in \V_D (\pi)$.
\bigskip

It is very well known that for $k=1$, requiring $\phi$ to be critical for any choice of $D$ identifies -- under fairly mild assumptions on $\vth$ -- a vector field $X_\vth$ in $M$, and critical sections are integral curves for $X_\vth$; in local coordinates all this takes the form of a system of ODEs which (with $t$ the coordinate on $B$) the functions $\phi (t)$ have to satisfy.

\medskip

In the standard higher dimensional generalization of this setting (i.e. the case $k > 1$), $I_D$ and $\de_V I_D$ are written as above, and requiring $(\de_V I_D) (\phi) = 0$ for all $V \in \V_D$ yields field equations, i.e. PDEs which $\phi (x^1,...,x^k)$ has to satisfy.

We show in this note that {\it a variational principle based on a $(n-2)$-form does also identify uniquely a vector field, i.e. a system of ODEs}. 

As mentioned above, such a variational problem naturally gives PDEs  (not an ODE), and a manifold (not a curve) as solution: we will show below that under suitable and rather general assumptions this manifold is necessarily an integral manifold for a one dimensional module (over the ring of functions $f : M \to \R$) of vector fields in $M$, i.e. identifies a vector field, unique up to multiplication for a scalar function.

In this way one does considerably enlarge the class of vector fields (dynamical systems) on $M$ defined by a variational principle; in particular, see section 5, any Liouville -- i.e. volume-preserving -- vector field admits such a characterization.

Here we want to focus on the mechanism at the basis of this fact; thus -- in order to avoid unessential difficulties -- we will work locally. We assume the reader is familiar with the language of (local) differential geometry.

In our discussion we will need some notions, briefly introduced below, from the theory of variational principles and of Cartan ideals.
We assume the reader is also familiar with the calculus of variations and just remind some basic definitions in section 1.1. On the other hand, Cartan ideals are maybe less standard and we provide the definition and results needed for our discussion in section 2.

\bigskip\noindent
{\bf Acknowledgements.} We would like to thank Enrico Massa for introducing us to Cartan ideals, and Giuseppe Marmo for discussions on the geometry of Liouville fields. Last but not least we most warmly thank an unknown referee of our paper \cite{GM} for kindly but firmly forcing us to develop our {\it ad hoc} approach to hyperhamiltonian vector fields into the general results exposed here. The work of GG was supported by {\it ``Fondazione CARIPLO per la ricerca scientifica''}.

\vfill\eject

\section{Variational principles and variational modules}

We start by recalling the general framework of variational problems, also in order to fix some notation needed for our discussion. We will then introduce the variational module associated to a variational problem, and study its annihilator.

Let $\pi : M \to B$ be a smooth bundle; we assume that $M$ is $n$-dimensional, and $B$ is a smooth manifold of dimension $k$, with $1 \le k < n$.

We denote by $\Ga (\pi )$ the set of smooth sections of the bundle $\pi : M \to B$, and by $\V (\pi)$ the set of vertical vector fields in $M$, i.e. of vector fields which are everywhere tangent to the fibers of the fibration $\pi$. If $D$ is a domain in $B$, we denote by $\V_D (\pi) \ss \V (\pi)$ the set of vertical vector fields which vanish on all of $\pi^{-1} (\pa D)$. All modules will be meant to be modules over $\Lambda^0 (M)$. We will use these notations for all bundles.

\subsection{Variational principles}

Consider a form $\vth \in \Lambda^k (M)$; then to any domain $D \ss B$ we associate a functional $I_D : \Ga (\pi ) \to \R$ by
$$ I_D (\phi) \ := \ \int_D \phi^* (\vth ) \ . \eqno(1) $$

Let $V\in \cal V(\pi)$ and $\gamma \in \Gamma (\pi)$; denote by $\psi_s$ the flow of $V$ on $M$. This induces a flow in $\Gamma$, and the flow of $\gamma$ is the one-parameter family of local sections  
$\widetilde{\psi}_s (\gamma):= \psi_s \circ\gamma$

The variation under $V$ of $I_D$ at $\phi \in \Ga (\pi )$ is defined as $$ (\de_V I_D ) (\phi) \ := \ {\d ~ \over \d s} \ \[ \int_D \, \( \widetilde{\psi}_s (\phi) \)^* (\vth) \]_{s=0} \ . \eqno(2) $$ 

The equation $(\de_V I_D ) (\phi ) = 0$ for all $V \in \V_D (\pi )$ [we write $\de I_D (\phi)$ for short] is the {\it variational principle on $\pi : M \to B$ defined by $\vth$}. With reference to the degree of $\vth$ (equal to the dimension of $B$), we say this is a {\it variational principle of degree $k$}.

A section $\phi \in \Gamma (\pi)$ is {\it critical for $I_D$} if and only if $(\de_V I_D ) (\phi) = 0 $ whenever $V \in {\cal V}_D (\pi)$.
It is well known (see e.g. \cite{Her}) that: 

\medskip\noindent
{\bf Proposition 1a.} {\it A section $\phi \in \Gamma (\pi)$ is critical for $I_D$ if and only if $ \phi^* ( V \interno \d \vth )  =  0$ for all  $V \in \V_D (\pi)$.}
\medskip

\Remark{1.}
Note that the forms $\vth $ and $\vth' = \vth + \a$ with $\a$ closed (locally, $\a = \d \b$) define the same $\d \vth = \d \vth'$ and thus the associated variational principles have the same critical sections $\phi$. $\odot$
\medskip

\Remark{2.} In our previous discussion, $k$ can be any number $0 < k < n$; however, the case $k = n-1$ is of little interest. 
Take local coordinates $(x^1 , ... , x^{n-1} ; z )$, say with $z$ the coordinate along fibers of $\pi : M \to B$. In this case $\d \vth = \rho (\xb,z) \Om$, with $\Om = \d x^1 \w ... \w \d x^{n-1} \w \d z$ and $\xb = (x^1,...,x^{n-1})$. Any vector field $V \in \V_D (\pi)$ is written as $V = f (\xb,z) \pa_z$ with $f$ vanishing on $\pi^{-1} (\pa D)$, and any $\phi \in \Ga (\pi)$ is described in these coordinates as $\phi = \{ (\xb,z) : z = u(\xb)\}$ for a suitable $u : \R^{n-1} \to \R$. With this, the pullback of $V \interno \d \vth$ is $ \phi^* [ (-1)^{n-1} f(\xb,z ) \rho (\xb , z) \, \d x^1 \w ... \w \d x^{n-1} ] \ = \ (-1)^{n-1} f(\xb, u (\xb)  ) \rho (\xb , u (\xb)) \ \d x^1 \w ... \w \d x^{n-1} $; hence the condition $\phi^* (V \interno \d \vth) = 0$ for all $V \in \V_D (\pi)$, i.e. with arbitrary $f$, just identifies the manifolds $z=u(\xb)$ on which $\rho (\xb , z) = 0$; these are algebraic -- and not differential -- equations in the $(\xb,z)$ variables. $\odot$

\Remark{3.} A form $\eta \in \Lambda (M)$ is said to be {\it basic for $\pi : M \to B$} if $V \interno \eta = 0$ for all $V \in \V (\pi)$. Note that the mechanism described in remark 2 will also act if $\vth$ is a basic form for $\pi : M \to B$, i.e. in this case the variational principle will not provide differential equations for $\phi$. $\odot$
\medskip

In view of remark 3, we say that the variational principle over $\pi : M \to B$ defined by $\vth$ is {\it proper} if $\vth$ is not basic for $\pi : M \to B$. We will always tacitly assume the variational principle we are considering is proper.

In view of remark 2, we say that a variational principle of degree $k = n-2$ is a {\it maximal degree} variational principle. 

The minimal degree case, $k=1$, corresponds to variational principles based on one-forms, which are well known to identify vector fields, i.e. to produce ordinary differential equations.

We will see below that (for any $k$) a critical section is also an integral manifold for a certain Cartan ideal; these critical manifolds have some interesting properties which make their determination simpler than what one would expect (basically by the method of characteristics). 

For maximal degree variational principles (i.e. for $k=n-2$) we will again be reduced to consider vector fields on $M$:   {\it maximal degree variational principles provide ordinary differential equations}.

\subsection{Variational modules}

Let $\pi:M \to B$ be the fiber bundle considered above, and $\vth \in \Lambda^k (M)$. 
Consider a basis $\{ V_1 , ... , V_r \} $ (here and below, $r=n-k$) of vertical vector fields, generating $\V (\pi )$ as a module. Then any $V \in \V (\pi )$ can be written as $V = \sum_{i=1}^r f^i (x) V_i $, and $V \in \V_D (\pi ) \ss \V (\pi)$ if and only if $f^i (x) = 0$ for all $x \in \pi^{-1} (\pa D)$ and for all $i=1,...,r$. 

Having chosen the $V_j$, we define the forms $\Psi_j \in \Lambda^k (M)$ as $ \Psi_j := V_j \interno \d \vth$ (for $j=1,...,r$); we denote the module generated by $\{ \Psi_1 , ... , \Psi_r \}$ as $\W (\pi,\vth )$. This is the {\it variational module} associated to the variational principle over $\pi : M \to B$ defined by $\vth$. 

\Remark{4.} Note that although $\Psi_j$ depend on the choice of a basis in $\V (\pi)$, the module $\W (\pi , \vth)$ does not depend on this choice, but only on the fibration $\pi$ and on the form $\vth$. $\odot$ 
\medskip

\Remark{5.} 
In general, $\W (\vth,\pi)$ is a $r$-dimensional module, where $r = n-k$ is the dimension of the module $\V (\pi)$; if $\W (\vth,\pi)$ has a lower dimension, we say it is degenerate. Note also that if $\vth$ is a basic form for $\pi : M \to B$, then the variational module $\W (\vth,\pi)$ is fully degenerate, i.e. one-dimensional. In the case $r=2$, $\W (\vth,\pi)$ is degenerate iff $\vth$ is basic for $\pi$. $\odot$ \medskip

We can restate proposition 1a as follows:

\medskip\noindent
{\bf Proposition 1b.} {\it A section $\phi \in \Gamma (\pi)$ is critical for $I_D$ if and only if  
$\phi^* (\Psi_j) = 0$ for all $j = 1,...,r$, i.e. iff $\phi^* (\Psi) = 0 $ for all $\Psi \in \W (\pi,\vth)$.} 
\medskip

Note that this condition is independent of $D$; on the other hand, $V \in \V_D (\pi)$ was needed for proposition 1a to hold (it eliminates boundary terms).

We denote as $\N (\eta)$ the annihilator of the form $\eta \in \Lambda (M)$; this is the module of vector fields $Y$ on $M$ such that $Y \interno \eta = 0$. 

\medskip\noindent
{\bf Lemma 1.} {\it Let $M$ be a smooth $n$-dimensional manifold, and $\eta \in \Lambda^{k+1} (M)$ be nonzero. 
Let $X;V_1,...,V_r$ be $r+1$ independent and nonzero vector fields on $M$, with $r=n-k$. 
Then $V_j \interno (X \interno \eta) = 0$ for all $j=1,...,r$ implies (and is thus equivalent to) $X \interno \eta = 0$.}

\medskip\noindent
{\bf Proof.} This is basically dimension counting. We use local coordinates $\{x^1 , ... , x^n \}$ in $M$ such that $V_j = \pa_j $. Then $V_j \interno (X \interno \eta) = 0$ for all $j=1,...,r$ implies that the form $\chi := X \interno \eta$ does not contain any $\d x^i$ factor for $i=1,...,r$. 
Note that $r=n-k$ and $\chi$ is a $k$-form: hence $\chi = F (x) \, \d x^{r+1} \w ... \w \d x^n$.
Recall now that $X$ was independent of $V_1,...,V_r$ and that $\chi = X \interno \eta$: these facts, and the form of $\chi$ given above, are in contradiction unless $\chi = 0$. $\triangle$

\medskip\noindent
{\bf Lemma 2.} {\it Let $\pi:M \to B$ be a smooth fiber bundle with $k$-dimensional base manifold $B$; let $\vth \in \Lambda^k (M)$ be nonzero and non basic for $\pi : M \to B$. Then a vector field $X \not\in \V (\pi) $ satisfies $X \interno \W (\vth,\pi) = 0$ iff $X \in \N (\d \vth)$.}

\medskip\noindent
{\bf Proof.} The expression $X \interno \W (\vth,\pi) = 0$ means that  $X \interno (V \interno \d \vth )) = 0$ $\forall V \in \V (\pi)$); for this it suffices that the relation is satisfied for vectors $V_1,...,V_r$ generating $\V (\pi)$. Obviously, $X \in \N (\d \vth)$ implies that $X \interno (V \interno \d \vth )) = 0$ for any (not only vertical) vector field $V$. Taking any generating set $\{ V_1 , ... , V_r \}$ for $\V (\pi)$ and $\eta = \d \vth$, lemma 1 proves the statement. Note that there we assumed $X$ is independent of a generating set $V_1,...V_r$ for $\V (\pi)$, i.e. this applies only if $X \not\in \V (\pi)$.  $\triangle$
\medskip

This lemma shows, in other words, that the set of vector fields which are transversal to fibers of $\pi$ and annihilate $\W (\vth,\pi)$ coincides with the set of vector fields in $\N (\d \vth)$ which are not vertical. 

We stress that there could be, in general, vertical vector fields which annihilate $\W (\vth,\pi)$; these are not necessarily in $\N (\d \vth)$. E.g., with the notation introduced above, consider the form $\eta = \beta \w \zeta$ where $\beta$ is a vertical one-form, and $\zeta$ is basic (see remark 3). Then, for any $\vth$ such that $\d \vth = \eta$ the vector field $X \in \V (\pi)$ such that $X \interno \beta = 1$ is in $\W (\vth,\pi)$ but not in $\N (\d \vth)$.

We also stress that for a generic nonzero $\eta$, we are not guaranteed that $\N (\eta ) \not= \{ 0 \}$, nor that there are nonzero independent vectors as required by lemma 1. Moreover, the rank of $\N_x (\eta ) := \{ \xi \in \T_x M : \eta_x (\xi)=0 \}$ could be different at different points $x \in M$. These problems are not present when $\eta \in \Lambda^{n-1} (M)$, as discussed below.

\subsection{The maximal degree case}

Let us now consider the case $k = n-2$, so that $\d \vth \in \Lambda^{n-1} (M)$. In this case lemmas 1 and 2 continue to hold, and moreover $\N (\d \vth)$ is always one dimensional; this is actually true for the annihilator $\N (\eta)$ of any nonzero $\eta \in \Lambda^{n-1} (M)$. It is easy to give an explicit expression for vector fields in $\N (\eta)$; this will also show that $\N (\eta)$ is one dimensional. 

Introduce local coordinates $(x^1 , ... , x^n)$ in $M$, such that the vertical coordinates for the $\pi : M \to B$ fibration are $x^1$ and $x^2$. We will write $ \Om = \d x^1 \w ... \w \d x^n$.

We can write a generic $(n-1)$-form $\eta$ and a vector field $X$ in $M$ as  
$$ \eta \ =  \ A^\mu \ (\pa_\mu \interno \Omega) \ \ , \ \  
X \ = \ f^\sigma \partial_\sigma \ , \eqno(3)$$
where the indices $\mu$ and $\sigma$ run from $1$ to $n$. If $\d \eta = 0$, as for $\eta = \d \vth$, then $\pa_\mu A^\mu = 0$.

It is then immediate to see that $X \interno \eta = 0$ if and only if 
$  f^\a A^\b = f^\b A^\a$ $\forall \a , \b = 1,...,n$; this of course entails 
$$f^\mu (x) \ = \ F (x) \, A^\mu (x) \ \ \ (F \in \Lambda^0 (M)) \ . \eqno(4) $$ 
We have thus at once the

\medskip\noindent
{\bf Lemma 3.} {\it If in the local coordinates $x^\mu$ the form $\d \vth$ is written as $\d \vth = A^\mu (\pa_\mu \interno \Om )$, then $\N (\d \vth)$ is a one dimensional module, generated by the vector field $X$ written in these coordinates as $X = A^\mu \pa_\mu$.} \medskip

In the following we will be interested in the case where one of the $A^\mu$ corresponding to horizontal coordinates, say $A^n$ for definiteness, never vanishes: $A^n (x) \not= 0$ $\forall x \in M$. 
Then the equations $f^\a A^\b = f^\b A^\a$ with $\a = n$ yield at once $ f^\b = ( A^\b / A^n ) f^n$. Note that all equations $f^\a A^\b = f^\b A^\a$, whatever $\a,\b$, are now automatically satisfied.

In this case, as $f^n$ is also nowhere vanishing (unless $X$ is identically zero), we can normalize $X$ requiring $X \interno \d x^n = 1$, which of course just means $f^n = 1$.

\section{Cartan ideals}

Variational principles can be formulated in terms of {\it Cartan ideals}, i.e. ideals of differential forms. In this section we will first recall some basic notions from the theory of Cartan ideals; the reader is referred to \cite{Car} for further detail, and all results quoted here can also be found in \cite{Arn,Bry,Olv}. 

A very readable account of Cartan's theory \cite{Car} in modern language is given in the final chapters of \cite{Olv}, and further developements are discussed in \cite{Bry}. See also \cite{Arn,God} for the use of Cartan's ideals in the study of PDEs and in analytical mechanics (including standard variational formulation of the latter). The relation between Cartan ideals and variational problem is studied in great detail, for $B$ one dimensional, in \cite{Gri}. The geometry of PDEs is naturally discussed in terms of Cartan ideals, see  e.g. \cite{AVL,Arn,Bry,Car,Olv}.
\medskip

We assume again that $M$ is a smooth $n$-dimensional manifold.

We say that $\J \ss \Lambda (M)$ is a {\it Cartan ideal} iff {\bf (i)} it is an ideal in $\Lambda (M)$ under exterior product, and {\bf (ii)} $\J_k := \J \cap \Lambda^k (M)$ is a module over $\Lambda^0 (M)$ for all $k =0,...,n$. In other words, $(i)$ for all $\eta \in \J$, $\psi \in \Lambda (M)$, $\eta \w \psi \in \J$; and $(ii)$ for  all $\beta_i \in \J_k$, $f_i \in \Lambda^0 (M)$ ($i=1,2$), $f_1 \beta_1 + f_2 \beta_2 \in \J_k$ (for all $k=0,...,n$).

Let $i : S \to M$ be a smooth submanifold of $M$ (here and below $i$ is the canonical inclusion); $S$ is said to be an {\rm integral manifold} of the Cartan ideal $\J$ iff $i^* (\eta ) = 0$ for all $\eta \in \J$. In other words, $S \ss M$ is an integral manifold of $\J$ iff all $\eta \in \J$ vanish on $S$.

The Cartan ideal $\J$ is said to be {\it generated} by the forms $\{ \eta^{(\a)} , \a= 1,...,r \}$ (with $\eta^{(\a)} \in \J$) if each $\zeta \in \J$ can be written as $\zeta = \sum_\a \rho_{(\a)} \w \eta^{(\a)}$ for a suitable choice of $\rho_{(\a)} \in \Lambda (M)$, $\a = 1,...,r$.

\medskip\noindent
{\bf Proposition 2.} {\it If $\J$ is generated by $\{ \eta^{(\a)} , \a= 1,...,r \}$, then $i:S \to M$ is an integral manifold for $\J$ iff $i^* (\eta^{(\a)}) = 0$ for all $\a = 1,...,r$.}
\medskip

The Cartan ideal $\J$ is said to be {\it closed} if it is closed under exterior differentiation, i.e. if $\d \eta \in \J$ for all $\eta \in \J$. In this case one also says that $\J$ is a {\it differential ideal}.

If the Cartan ideal $\J$ is generated by $\{ \eta^{(\a)} , \a= 1,...,r \}$, it can always be completed to a differential ideal by adding the  $\d \eta^{(\a)} \not\in \J$ to the system of generators. We denote by $\widehat{\J} $ the completion of the ideal $\J$ obtained in this way; obviously $\J \sse \widehat{\J}$, the equality corresponding to the case where $\J$ is closed.

Note that if $\eta$ vanishes on $S$, the same is true of $\d \eta$; thus, the integral manifolds of $\J$ and of $\widehat{\J}$ coincide.
In Cartan's words, ``La recherche des solutions d'un syst\`eme diff\'erentiel peut toujours \'etre ramen\'ee \`a la recherche des solutions d'un syst\`eme diff\'erentiel ferm\'e '' (see \cite{Car}, p. 52).

We will always assume that $\J$ does not include 0-forms; by the previous remark, this is not actually a limitation (but simplifies discussions).
\bigskip

Given a Cartan ideal $\J$, we associate to any point $x \in M$ the subspace $D_x (\J) \ss \T_x M$ defined by
$ D_x (\J) := \{ \xi \in \T_x M  : \, \xi \interno \J_x \ss \J_x \}$. 

If $D_x (\J)$ has constant dimension, the Cartan ideal $\J$ is said to be {\it non singular}, and the distribution $D (\J) = \{ D_x (\J) , x \in M\}$ is its {\it characteristic distribution}; any vector field $X \in D (\J)$ (by this we mean that $X (x) \in D_x (\J )$ at all points $x \in M$) is said to be a {\it characteristic field} for $\J$. 

\Remark{6.} 
Note that if all the generators $\eta^{(\a)}$ of $\J$ are of the same degree $k$,  then all forms in $\J$ are of degree not smaller than $k$, and $\J_m = \{ 0 \}$ for $m < k$. If $\J_m = \{ 0 \} $ for $m<k$, then $X \in D (\J)$ satisfies $X \interno \zeta = 0$ for all $\zeta \in \J_k$, and in particular $X \in D (\J)$ iff $X \interno \eta^{(\a)} = 0$. Indeed by definition any $\zeta \in \J$ is written as $\zeta = \rho_{(\a)} \w \eta^{(\a)}$, and $X \interno \zeta = \s_{(\a)} \w \eta^{(\a)}$ with $\s_{(\a)} = X \interno \rho_{(\a)}$. $\odot$ \medskip

An {\it integral manifold} for a distribution $D$ on $M$ is a submanifold $i:N \to M$ such that $i_* (\T_x N ) \ss D_{i (x)}$ for all $x \in N$. In other words, any vector field tangent to $N$ is in  $D$ (the converse is in general not true).

\Remark{7.} It is easy to see that integral manifolds of $D(\J)$ are always integral manifolds of $\J$. It is maybe worth stressing that the converse is in general not true: just consider a symplectic form $\om$ in $\R^{2n}$ and the ideal $\J$ generated by $\om$; the only vector field satisfying $X \interno \J \ss \J$, i.e. $X \interno \om = 0$ (see remark 6), is $X \equiv 0$; on the other hand, lagrangian manifolds are integral manifolds of $\J$. $\odot$
\medskip

The $d$-dimensional distribution $D$ on $M$ is said to be {\it completely integrable} if through each point $x \in M$ passes a $d$-dimensional integral manifold of $D$. In this case, the $d$-dimensional integral submanifolds are also said to be the {\it Cauchy characteristics} for $D$.

\medskip\noindent
{\bf Proposition 3.} {\it If $\J$ is a closed nonsingular differential Cartan ideal, then $D (\J)$ is completely integrable.}
\medskip

It should be stressed that the Cauchy characteristics of an integrable  $d$-dimensional distribution $D$ provide a foliation of $M$ by $d$-dimensional submanifolds \cite{Nar}. 

Thus if $\J$ is a closed nonsingular Cartan ideal with $d$-dimensional characteristic distribution $D (\J)$, then $\J$ always has $d$-dimensional integral manifolds, and $M$ is foliated by these. In the following we will deal in particular with the case $d=1$.

The following theorem (proposition 4) is most useful in performing computations with Cartan ideals; it appears in different forms in \cite{Bry,Car,Olv}. Here we will use an immediate consequence of it, i.e.  proposition 5; see section 45 of \cite{Car}.

\medskip\noindent
{\bf Proposition 4.} {\it Let $\J$ be a nonsingular differential Cartan ideal, and let its characteristic distribution $\D (\J)$ be $p$-dimensional. Then in a neighbourhood of any point $x \in M$ we can choose local coordinates $(x^1 , ... , x^p ; y^1 , ... , y^{n-p})$ such that $\J$ admits a system of generators $\{ \th_1 , ... , \th_r \}$ with the property that, locally around $x$, the $\th$ and $\d \th$ do not involve the variables $x^j$ nor the one forms $\d x^j$.}
\medskip

The local coordinates whose existence is guaranteed by this theorem will be called {\it Cartan canonical coordinates}; if we consider locally a fibration of $M$ over $\R^p$ for which the $x^i$ are horizontal and the $y^j$ are vertical coordinates, $\D (\J)$ spans horizontal planes identified as $y^j = const$, $j=1,...,n-p$.

\medskip\noindent
{\bf Proposition 5.} {\it Let $X$ be a characteristic vector field for $\J$, and let $i: S \to M$ be a $q$-dimensional integral manifold of $\J$. Assume that $X$ is nowhere tangent to $i (S)$. Let $\Phi_t $ be the local one-parameter group of diffeomorphisms generated by $X$. The $(q+1)$-dimensional manifold $\Phi : (- \varepsilon , \varepsilon ) \times S \to M$ defined by $\Phi (\tau , x) = \Phi_\tau (x)$ is an integral manifold of $\J$.}

\section{Cartan ideals and variational principles}

We have seen in section 1 that the variational principle over $\pi : M \to B$ defined by $\vth$ defines the module $\W (\pi,\vth)$. 

We will now consider the Cartan ideal $\J$ generated by $\W (\pi,\vth)$; by this we mean the ideal generated by a set of generators of $\W (\pi,\vth)$, i.e. of generators $V_j$ for $\V (\pi)$. Note that by remark 4 this does not depend on the choice of the $V_j$. 

\medskip\noindent 
{\bf Definition.} The Cartan ideal  $\J (\vth,\pi)$  generated by $\W (\vth,\pi)$ is the {\it Cartan ideal associated to the variational principle on $\pi$ defined by $\vth$.}
\medskip

Note that if $(\d \vth)_{x_0} = 0$ at some point $x_0 \in M$, then $\Psi_j = \pa_j \interno \d \vth$ also vanish at that point, and $D_{x_0} (\J) = \T_{x_0} M$. Thus in order to have a nonsingular $\J (\vth , \pi)$, we have to require that $\d \vth$ is nowhere zero.

We can characterize critical sections of the variational principle on $\pi : M \to B$ defined by $\vth$ by noting that: {\it the critical sections of the variational principle on $\pi : M \to B$ based on $\vth$, with $\d \vth$ nowhere vanishing on $M$, are integral manifolds of the Cartan ideal $\J (\vth,\pi)$.}
We can therefore rephrase proposition 1b (which was a restatement of proposition 1a) in terms of Cartan ideals. 

\medskip\noindent
{\bf Proposition 1c.} {\it A section $\phi \in \Gamma (\pi)$ is critical for the variational principle on $\pi : M \to B$ defined by $\vth$ if and only if $\phi$ is an integral manifold of the Cartan ideal $\J (\vth,\pi)$.}
\medskip

This proposition justifies calling $\J (\vth,\pi)$ the Cartan ideal associated to the variational principle $\de I_D  = 0$: indeed, it implies that in order to study (critical sections for) the variational principle $(\de I_D)(\phi) = 0$, we can just study (integral manifolds of) the Cartan ideal $\J(\vth,\pi )$.

More precisely, we have to study integral manifolds of $\J (\vth,\pi)$ that are sections of $\pi : M \to B$; this means in particular that they are of dimension $k$ and everywhere transversal to fibers of the bundle $\pi : M \to B$. 

We have thus completely characterized critical sections $\phi$ for a variational principle as sections which are integral manifolds for the associated Cartan ideal.
This approach will be particularly useful in the case of maximal degree variational principles.

\subsection{General reduction}

Consider the variational principle on $\pi : M \to B$ defined by $\vth \in \Lambda^k (M)$; assume $\J := \J (\vth,\pi)$ is nonsingular, and $\D := D[\J (\vth,\pi)]$ is $q$-dimensional.

The result of proposition 5 can be applied to reduce the problem of determining critical section of a variational principle, i.e. $k$-dimensional integral manifolds of $\J (\vth,\pi)$, down to that of determining $(k-q)$-dimensional one satisfying suitable transversality conditions.

This rests on the possibly of applying several times proposition 5.
We say that the submanifold $M_0 \ss M$ is {\it non characteristic} for $\J$ if $T_x M_0 \cap [D (\J)]_x = \{ 0 \}$ for all $x \in M_0$. Then a local integral manifold for $\J$ is specified by assigning a manifold $M_0$ which is integral and non characteristic for $\J$, and "pulling" it along the characteristic distribution $\D$. 

In a less pictorial way, we build a local integral manifold for $\J$ as a local bundle over $M_0$, with fibers corresponding to integral manifolds for $\D$ (see proposition 3 and the remark after it); note this only uses the Frobenius integrability of $\D$ \cite{Arn,Car}.

Such a general reduction is {\it not} always possible; actually when the fibers of $\pi : M \to B$ have dimension greater than two it is generally impossible to perform it, as we now briefly discuss.

When looking for integral manifolds of $\J$ which are sections of $\pi : M \to B$, this general reduction would require to consider the subset $\D_\pi \sse \D$ which is transversal to fibers of $\pi : M \to B$, and extend integral manifolds of $\J$ over a submanifold $B_0 \ss B$ of codimension equal to the dimension of $\D_\pi$ to a local critical section. 

We stress that one should require several additional conditions for the reduction procedure to be viable: the dimension of $\D_\pi$ can vary even if that of $\D$ is constant; moreover, the involutivity of $\D$ does not imply, in general, involutivity and hence integrability of $\D_\pi$.

In practice, this means that this approach can be applied to the construction of critical sections only if $D [\J (\vth,\pi)]$ is transversal to the fibers of $\pi : M \to B$; this can be imposed by suitable nondegeneracy conditions on $\vth$ or equivalently on $\W (\vth,\pi)$.

Note also that $\N (\d \vth )$ (and thus the "useful" part of $D (\J)$, see lemma 2) is in general empty when $\vth$ does not have degree $k = n-2$. 
In this case, of course, we miss the main ingredient of the reduction procedure; the previous discussion shows that even when $\N (\d \vth ) \not= \emptyset$, we have to require nontrivial extra conditions.

\Remark{8.}
The above discussion can be reinterpreted in terms of the Cartan canonical coordinates introduced above; these define, indeed, a natural fibration $\kappa : M \to L$ over a $p$-dimensional manifold $L$, spanned (in the notation of proposition 4) by the coordinates $x^1 , ... , x^p$.
Thus, once $\vth$ and $\pi$ -- and thus $\J := \J (\vth,\pi)$ -- are fixed, we have two different local fibrations for $M$: the one dictated by the variational principle, i.e. $\pi : M \to B$; and the one corresponding to Cartan canonical coordinates for $\J$, i.e.  $\kappa : M \to L$. The latter is such that $D (\J)$ is transversal to fibers $\kappa^{-1} (\ell)$ for all $\ell \in L$, but in order to apply the reduction procedure we need that $D(\J)$ is transversal to fibers $\pi^{-1} (b)$ for all $b \in B$; this condition is in general not satisfied. $\odot$
\medskip

We stress that in the maximal degree case, the requirement that $\vth$ is non basic suffices to guarantee transversality, hence that this approach can be effectively used in the search for critical sections.

\subsection{Maximal degree variational principles}

We can now apply the previous general discussion to the study of maximal degree variational principles; we will freely use the concepts and notations introduced in section 1 (with $k = n-2$). 

\medskip\noindent
{\bf Theorem 1.} {\it Let $\pi : M \to B$ be a smooth fiber bundle of dimension $n$ with base manifold $B$ of dimension $k= n-2$; let $\vth \in \Lambda^k (M)$ be non basic for this fibration, and such that $\eta := \d \vth$ is nowhere zero on $M$. Then the Cartan ideal $\J (\vth,\pi)$ is nonsingular and admits a one-dimensional characteristic distribution $D[\J (\vth,\pi )]$; this coincides with $\N (\d \vth )$.}

\medskip\noindent
{\bf Proof.} First of all we notice that in this case $\N (\d \vth)$ is one dimensional, see lemma 3; we are also guaranteed this is not vertical, as $\vth$ is assumed to be non basic. 

Recall now that $\J (\vth,\pi)$ is generated by $\Psi_1,\Psi_2$ with $\Psi_j = V_j \interno \d \vth$, where $V_1,V_2$ generate $\V (\pi)$. As $\Psi_1, \Psi_2$ are both $k$-forms, $(X \interno \Psi_j) \in \J$ is equivalent to $X \interno \Psi_j = 0$ (see remark 6), i.e. to $X \interno \W (\vth,\pi) = 0$; thus this theorem is merely a restatement of lemma 2, for the case $k=n-2$, in the language of Cartan ideals. 

Note that lemma 3 and the condition $\eta \not= 0 $ at all points $x \in M$ imply that the linear subspace $D_x$  of vectors $\xi \in \T_x M$ satisfying $\xi \interno \eta = 0$ has constant dimension, i.e. $\J (\vth,\pi)$ is nonsingular. $\triangle$
\medskip

Let us now consider a smooth submanifold $B_0 \ss B$ of codimension one in $B$, and let $\pi_0 : M_0 \to B_0$ be the associated subbundle of $\pi : M \to B$; this is defined by $M_0 = \pi^{-1} (B_0) \ss M$, with $\pi_0$ the restriction of $\pi$ to $M_0$. We denote, with our standard notation, sections of this subbundle as $\Ga (\pi_0 )$.

\medskip\noindent
{\bf Theorem 2.} {\it Let $B_0 \ss B$ be a smooth submanifold of codimension one in $B$, and $\pi_0 : \pi^{-1}(B_0) \to B_0$ the associated subbundle of $\pi: M \to B$. Let $\phi_0 \in \Ga (\pi_0)$, seen as a submanifold of $M$, be an integral manifold for the Cartan ideal $\J (\vth,\pi)$, nowhere tangent to integral manifolds of $D[\J (\vth,\pi)]$. 

Then the critical local sections for the maximal degree variational principle on $\pi$ defined by $\vth$ can be built by pulling $\phi_0$ along integral curves of $D [\J (\vth,\pi)]$.}

\medskip\noindent
{\bf Proof.} This follows immediately from proposition 5 and lemma 3, which guarantees $D [\J (\vth,\pi)]$ is transversal to fibers of $\pi$ in the maximal degree case. Note that here we are only using the existence of local solutions to flows defined by smooth vector fields. $\triangle$ \medskip

Vector fields in $D [\J (\vth,\pi)]$ differ only by a nonzero function, and can thus be uniquely determined by a normalization prescription (e.g. setting $F(x) \equiv 1$ as in lemma 3). Thus, as announced, a maximal degree proper variational principle over $\pi : M \to B$ determines a unique vector field in $M$.

One could as well consider the inverse problem: given a vector field $X$ on $M$, is it possible to characterize it (up to normalization) in terms of a maximal degree variational principle ? Our previous discussion shows that we have the following:

\medskip\noindent
{\bf Theorem 3.} {\it Let $M$ be a smooth $n$-dimensional manifold, and $X$ a vector field on $M$. Assume there is an exact form $\eta = \d \vth \in \Lambda^{n-1} (M)$ such that: (i) $X \in \N (\eta)$ (ii) $\eta$ is nowhere vanishing, (iii) $\eta$ is not basic for the fibration $\pi : M \to B$ over a $(n-2)$-dimensional manifold $B \ss M$. Then $X$ generates the characteristic distribution of the Cartan ideal associated to the (maximal degree) variational principle on $\pi : M \to B$ defined by $\vth$.}

\section{Dynamics}

In our discussion we considered all directions in $M$ on the same footing. However in many applications we are interested in a slightly different situation, i.e. $M = P \times T$, where $P$ is a smooth manifold of dimension $p=n-1$ (the {\it phase space}), and $T$ is the real line $\R$ corresponding to the time coordinate $t$ (we also say that $M$ is the {\it extended phase space}). Similarly, we require one of the variables in the basis $B$ (these are the independent variables for the resulting PDEs) to be the time variable, i.e. $B = Q \times T$. 

In this case it is obvious that $B$ is itself a bundle $\beta : B \to T$, with fibers $\beta^{-1} (t) = Q$, and we have a double fibration
$$ M \mapright{\pi} B \mapright{\beta} T \ \ ; \ \ M \mapright{\tau} T  \ \ \ \ \ (\tau \equiv \beta \circ \pi ) \ . $$

Recall that if $\vth \in \Lambda^{n-2} (M)$, then by lemma 3, $\N (\d \vth)$ is one dimensional. We say that $\vth \in \Lambda^{n-2} (M)$ is transversal for $\tau$ if $\N (\d \vth)$ is everywhere transversal to fibers $\tau^{-1} (t)$.

\medskip\noindent
{\bf Lemma 4.} {\it In the conditions of theorem 1, let the base space $B$ of the bundle $\pi : M \to B$ be itself a bundle $\beta : B \to T = \R$, and consider the bundle $\tau : M \to T$ with $\tau = \beta \circ \pi$. Let $\vth \in \Lambda^{n-2} (M)$ be transversal for $\tau$. Then $\N (\d \vth) = D [\J(\vth,\pi)]$ is generated by a vector field of the form $X_0 = \pa_t + Y$, with $Y \in \V (\tau)$.}

\medskip\noindent
{\bf Proof.} Choose local coordinates $(x^0 , ... , x^p )$ in $M = P \times T$, with $(x^1 , ... , x^p )$ coordinates on $P$ and $T$ spanned by the $x^0 \equiv t$ coordinate.

By lemma 2 and lemma 3, any $X \in \N (\d \vth )$ is written as 
$X = \ga (x) [ f^\mu (x)  \pa_\mu ]$ with $\ga : M \to \R$ an arbitrary function and $f^\mu : M \to \R$ some given functions -- determined above -- satisfying $| f (x) |^2 := \sum_\mu [f^\mu (x)]^2 \not= 0$ for all $x \in M$. 

The transversality of $D [\J(\vth,\pi)]$ to fibers of $\tau : M \to T$  implies that actually $f^0 (x) \not= 0$ for all $x \in M$; thus we can choose $\ga = (f^0)^{-1}$ and obtain (we write $x^0 = t$) $X_0 = \pa_t + \sum_{j=1}^p b^j (x) \pa_j $ with $b^j = f^j / f^0$). 

This is of the form announced in the statement and nowhere zero by construction; as we know that $D [\J(\vth,\pi)]$ is a one dimensional module, $X_0$ is a generator for it.   $\triangle$

\Remark{9.} The expression of $X_0$ built in the proof coincides with the one obtained in section 1.3 (here $t$ takes the place of $x^n$). $\odot$
\medskip

Lemma 4 means that  -- provided $\vth$ is transversal for $\tau$  -- the variational principle $\de I_D (\phi) = 0$ (see 1.1) defines a unique normalized dynamical vector field in $P$. Indeed, the normalized generator $X_0$ (satisfying $X_0 \interno \d t = 1$) of $D (\J)$ can be written as $X_0 = \pa_t + Y_0$ and, since $Y_0 \in \V (\tau )$, this $Y_0$ is the required dynamical vector field in $P$.

\section{Variational structure of Liouville dynamics}

In this section we show that any Liouville (i.e. volume-preserving) vector field can be identified by a suitable maximal degree variational principle.

Let $P$ be a smooth $p$-dimensional manifold, and $\Om$ a volume form on it, i.e. a nowhere vanishing form $\Om \in \Lambda^p (P)$. A vector field $X$ on $P$ is said to be a Liouville field (with respect to $\Om$) if $L_X (\Om ) = 0$; this is locally equivalent to the condition that $\exists \ga \in \Lambda^{p-2} (P)$ such that 
$$ X \interno \Om \ = \ \d \ga \ .  \eqno(5) $$
For a discussion of the geometry of Liouville dynamics, see \cite{Mar}.

In this section we discuss the variational structure of Liouville  vector fields, in terms of a higher degree variational principle.

\medskip\noindent 
{\bf Remark 10.} Note that for $X$ and $\Om$ given, $\ga$ is not uniquely defined by (5); on the other hand, for $\ga$ and $\Om$ given, $X$ is uniquely defined by (5). $\odot$
\medskip

It is well known that if $P$ is a symplectic manifold, any Hamiltonian vector field on $P$ is also Liouville; the converse is not true. A rather popular case of non-hamiltonian Liouville dynamics is provided by Nambu dynamics \cite{Nam}; another recently introduced generalization of hamiltonian dynamics (still Liouville) is the hyperhamiltonian one \cite{GM,MT}. These relevant special cases are briefly discussed below. 

\subsection{General Liouville dynamics}

We consider the smooth $n$-dimensional manifold $M = P \times \R$, which of course is a bundle $\tau : M \to T \equiv \R$; we denote by $t$ the coordinate along the $T$ factor, and obviously $P$ is a smooth manifold of dimension $p= n-1$. 

We want to consider $M$ as a bundle $\pi : M \to B$ over a smooth $k$-dimensional manifold $B = Q \times T$, with $Q \ss P$ a smooth manifold of dimension $q=p-2$.

We denote by $\Om$ the volume form in $P$. Choose a form $\s \in \Lambda^{p-1} (P)$ which satisfies $\d \s = \Om$. Then, if $X$ satisfies (5), we define 
$$ \vth \ := \s \, + \, (-1)^p \, \ga \w \d t \ \in \ \Lambda^{n-2} (M) \ . \eqno(6) $$

\Remark{11.} Note that $\d \vth = \d \s \pm \d \ga \w \d t$ is nowhere zero: indeed $\d \s = \Om$ does not contain the $\d t$ factor, and thus $\d \vth$ can vanish only if both terms do; but by definition $\Om$ is nowhere zero. Note also that $\s$ is surely not basic for $\pi : M \to B$, whatever the choice of $B$; indeed it does not contain $\d t$ and is a $p-1 = q+1$ form on $P$. $\odot$
\medskip

\medskip\noindent
{\bf Theorem 4.} {\it The Liouville dynamics $X$ on $P$ satisfying (5) is associated to a vector field $Z = \pa_t + X$ on $M = P \times T$ identified by the maximal degree variational principle on $\pi : M \to B$ defined by $\vth$ given in (6).}

\medskip\noindent 
{\bf Proof.} Remark 11 shows that the variational principle based on $\vth$ is well defined; due to the discussion of sections 3 and 4, it  yields a $Z$ satisfying $Z \interno \d \vth = 0$. We only have to prove that this is of the form $Z = \pa_t + X$, with $X$ a vector field on $P$ (i.e. such that $X \interno \d t = 0$) such that $X \interno \Om = \d \ga$.

As $\d \vth$ is a nowhere vanishing $n-1$ form on a $n$-dimensional manifold, the set $\N (\d \vth )$ is one dimensional. We just have to identify a (nowhere zero) vector field generating this annihilator;  imposing the normalization condition $Z \interno \d t = 1$ will determine a unique vector field from it.

Consider vector fields of the form $Z = \pa_t + X$ with $X \interno \d t = 0$, as suggested in the statement. By definition of $\vth$, and assuming (5) is satisfied, 
$$ \begin{array}{rl}
Z \interno \d \vth \ =& \ X \interno \Om \ + \ (-1)^{p} \, (Z  \interno \d \ga) \w \d t \ + \ (-1)^{(2p-1)} \, \d \ga \\
 =& \ \d \ga \ + \ (-1)^p \, (X \interno \d \ga) \w \d t \ - \ \d \ga \\
 =& \ (-1)^{p} \, (X \interno \d \ga ) \w \d t \ = \ 
(-1)^p \, [ X \interno (X \interno \Om)] \ = \ 0  \ . \end{array} $$
This completes the proof. $\triangle$

\subsection{Hamilton dynamics}

As well known, any hamiltonian vector field is also Liouville. Let us describe how this is identified by a maximal degree variational principle (beside the standard minimal degree variational principle based on the Poincar\'e-Cartan one-form).

Let $P$ be a smooth manifold of dimension $p=2m$, equipped with a symplectic form $\om$; we write $\zeta = (1/(m-1)!) (\om)^{(m-1)}$ (this is obviously an exterior power). Choose a form $\b \in \Lambda^1 (P)$ such that locally $\om = m \d \b$.

The smooth function $H : P \to \R$ defines the hamiltonian vector field $X$ by $X \interno \om = \d H$. On the other hand $\Om = (1/m!) (\om)^{m} = (1/m) \om \w \zeta$. Thus 
$$ X \interno \Om \ = \ (X \interno \om ) \w \zeta \ = \ \d H \w \zeta \ , $$
and $X$ satisfies (5) with $\ga = H \, \zeta$.  

It follows immediately that the corresponding maximal degree variational principle is based on the form $\vth \in \Lambda^{2m-1} (M \times \R)$ given by 
$$ \vth \ = \ \b \w \zeta \ + \ H \, \zeta \w \d t \ = \ (\b \, + \, H \, \d t ) \w \zeta \ ; $$
the Hamilton equations are readily recovered from this. 

\subsection{Nambu dynamics}

Nambu dynamics \cite{Nam} encountered a renewal of interest in recent years; see \cite{Mor} for a discussion of it in terms of forms and Cartan ideals.
It is well known that Nambu dynamics is also Liouville, and that in general it cannot be described in terms of a standard (i.e. degree one) variational principle.

An intrinsic definition of Nambu vector fields is as follows: consider a smooth $n$-dimensional manifold $P$ with volume form $\Om$. Then the vector field $X$ on $P$ is Nambu if there is a choice of $n-1$ smooth functions  $H_i : P \to \R$ ($i= 2,...,n$) such that 
$$ \d H_2 \w ... \w \d H_n  \ := \ \ \chi \ = \ X \interno \Om \ . $$

We have immediately that $\chi = \d \ga$ with e.g. 
$$ \ga \ = \ H_2 \, \d H_3 \w ... \w \d H_n \ . $$
We also write e.g. $\s = x^1 \d x^2 \w ... \w x^n$, which yields $\d \s = \Om$. (For both $\ga$ and $\s$ one could use a different permutation of indices).

With these, $\vth$ is readily recovered, see (6), and hence -- for any $X$ -- we have determined the maximal degree variational principle defining the Nambu vector field $X$.

\subsection{Hyperhamiltonian vector fields}

Another special class of Liouville vector fields is provided by {\it hyperhamiltonian} vector fields, generalizing Hamilton dynamics and  studied in \cite{GM,MT}; these are based on hyperkahler (rather than symplectic) structures.

In this case, one considers a riemannian manifold $(P,g)$ of dimension $p=4N$, equipped with three independent symplectic structures $\om_\a$ ($\a = 1,2,3$); to a triple of smooth functions $\h^\a : P \to \R$ one associates a triple of vector fields by (no sum on $\a$) $ X_\a  \interno  \om_\a  =  \d  \h^\a $. The {\it hyperhamiltonian vector field} $X$ on $P$ associated to the triple $\{ \h^\a \}$ is the sum of these, $ X := \sum_{\a=1}^3  X_\a $; it is trivial to check that the $X_\a$, and therefore $X$, are uniquely defined. Each $X_\a$ is obviously Liouville, and so is $X$.
 
On the $(p+1)$ dimensional manifold $M = P \times \R$ (denote by $t$ the coordinate on $\R$) the time evolution under $X$ is described by the vector field $Z = \pa_t + X$. 

Let $\s_\a$ be one-forms satisfying $\d \s_\a = \om_\a$, and $\z_\a$ the $(2N-1)$-th exterior power of $\om_\a$. Define (with $s=\pm 1$ taking care of orientation matters \cite{GM})
$$ \vth \ = \ \sum_{\a=1}^3 \ \s_\a \w \z_\a \ + \ (6 N s) \, \sum_{\a=1}^3 \ \h^\a \ \z_\a \w \d t \ .  $$

It is immediate to check that $\d \vth$ is nonsingular, and that 
$ Z \interno \d \vth = 0$, $Z \interno \d t = 1$.
It follows from our general discussion that the vector field $Z$ is also obtained by a maximal degree variational principles based on the form $\vth$.

\vfill\eject

\end{document}